\documentclass[a4paper]{elsarticle}
\usepackage{lineno,hyperref}

\usepackage[utf8]{inputenc}
\usepackage{bm}
\usepackage{amsmath}
\usepackage{appendix}
\usepackage{xcolor}
\usepackage{amssymb}
\usepackage{soul}

\usepackage{nomencl}
\makenomenclature
\journal{Journal of \LaTeX\ Templates}

\begin{document}

\begin{frontmatter}


\title{Predictive Model and Optimization of Micromixers Geometry using Gaussian Process with Uncertainty Quantification and Genetic Algorithm}


\author[1,2]{Daniela de Oliveira Maionchi\corref{mycorrespondingauthor}}
\cortext[mycorrespondingauthor]{Corresponding author}
\ead{dmaionchi@fisica.ufmt.br}
\author[1]{Neil Diogo Silva Coimbra}
\author[1]{Júnior Gonçalves da Silva}
\author[2]{Fabio Pereira dos Santos}

\address[1]{Instituto de F\'isica , Universidade Federal de Mato Grosso - UFMT, 78060-900, Cuiab\'a-MT, Brazil}
\address[2]{Escola de Qu\'imica, Universidade Federal do Rio de Janeiro - UFRJ, 21941-909, Rio de Janeiro-RJ, Brazil}

\begin{abstract}

Microfluidic devices are gaining attention for their small size and ability to handle tiny fluid volumes. Mixing fluids efficiently at this scale, known as micromixing, is crucial. This article builds upon previous research by introducing a novel optimization approach in microfluidics, combining Computational Fluid Dynamics (CFD) with Machine Learning (ML) techniques.  The research focuses on improving global optimization while reducing computational expenses. It draws inspiration from a Y-type micromixer, initially featuring cylindrical grooves on the main channel's surface and internal obstructions. Simulations, conducted using OpenFOAM software, evaluate the impact of circular obstructions on mixing percentage and pressure drop, considering variations in obstruction diameter and offset. A Gaussian Process (GP) was utilized to model the data, providing model uncertainty. Thus, this study optimizes geometries by using  genetic algorithm (GA) and least-square optimization based on the reduced order model provided by GP. Results align with previous research, showing that medium-sized obstructions (137 mm diameter, 10 mm offset) near the channel wall are optimal. This approach not only provides efficient microfluidic optimization with uncertainty quantification but also highlights the effectiveness of combining CFD and ML techniques in achieving desired outcomes. 
\end{abstract}

\end{frontmatter}

\begin{table}[ht]
	\centering
	\caption{Nomenclature.}
	\label{symbols}
	\small
	\begin{tabular}{l l |l l}
	$C$ & concentration &    $Re$ & Reynolds number   \\
    $CG$ & cylindrical grooves & $Sc$ & Schmidt number \\
	$d$    & channel depth   & $u$ & velocity \\
	$L$   & channel lenght     & $W$   & channel width \\
    $\rho$ & density 	 & $\Delta P$ & pressure drop \\
    $\sigma$ & standard deviation  &  $\varphi$ & mixing percentage \\	
	$OD$  & obstruction diameter & $\mu$ & viscosity  \\
    $OF$ & offset & $\gamma$ & diffusion coefficient\\
	\end{tabular}
\end{table}

\section{Introduction}

Optimizing microfluidic device designs traditionally relies heavily on experimentation and Computational Fluid Dynamics (CFD) simulations. However, these methods have limitations.  Creating datasets through experimentation can be expensive and time-consuming, requiring the fabrication and testing of numerous device geometries. Similarly, while CFD offers valuable insights, it can be computationally expensive, especially when running a large number of simulations to explore the design space. Here, Machine Learning (ML) offers a potent alternative. It has already become an integral part of the quality by design strategy and has been introduced to fields such as Additive Manufacturing and Microfluidics. ML enables the prediction and generation of learning patterns for precise fabrication \cite{dedeloudi2023}.
By leveraging data-driven approaches, ML can significantly reduce the number of simulations needed for optimization compared to traditional methods.  While CFD-ML integration is gaining attraction in other fields, its application in microfluidics remains relatively unexplored. This research aims to bridge this gap by utilizing ML to optimize microfluidic device designs, reducing both experimental and computational costs. 

Among the research focusing on passive micromixers, \cite{mustafa2023} investigated a hybrid T-model featuring various combinations of twist and bend angles across different Reynolds numbers (Re). Micromixer optimization was achieved through computational fluid dynamics (CFD) and multi-objective optimization (MFO). The findings, consistent with prior research \cite{sinha2022, prakash2021}, indicated a monotonic increase in mixing efficiency with higher twist and bend angles. Additionally, the results demonstrated that both pressure drop ($\Delta$P) and mixing energy costs rise with increasing twist and bend angles, particularly as Reynolds number increases. Traditional optimization methods, like CFD simulations, often require many iterations for global optimization, which can take months. However, leveraging machine learning techniques could streamline this process, reducing the required time to as little as 35 days~\cite{maionchi2022}.

Similar studies were conducted by \cite{nikpour2022} and \cite{yang2022}. Both studies investigate a Y-type passive planar micromixer. The former employs triangular, diamond, and circular baffles distributed along the micromixer, while the latter utilizes inclined rectangles with defined length and angles along the micromixer walls. Both used the response surface method (RSM) method to model the mixing index and pressure drop and the designs were ranked among the members of the Pareto front by multi-objective optimization. In \citep{sun2022}, micromixer designs that consist of Tesla/hexagonal chambers and $\Omega$-shaped obstacles or combined with straight-shaped obstacles are analysed. The authors obtain an optimal geometry with mixing efficiency greater than 78\% and maximum pressure drop of 61.5kPa, which is beneficial for the on-chip integration.

A variation of the aforementioned works is presented in \cite{yang2023}, where it proposes to determine the values of operational parameters of a microfluidic concentration gradient generator to generate user-specific concentration gradients of biomolecules. For such complex optimization problem, a robust multi-fidelity machine learning model named neural-physics multi-fidelity model was adopted. A physics-based component model based on Fourier series expansion and a CFD model  utilized to simulate fluid flow and chemical species mixing and generate data for neural network construction. In order to accelerate convergence during optimization, this kind of calculation needed to be performed on a GPU platform to utilize its massively parallel processing threads.

A previous study by \cite{maionchi2022} investigated the geometry proposed in \cite{wang2012} and utilized neural networks to fit the generated data, incorporating variations in the diameter (OD) and offset (OF) of circular obstacles. The study employed a multiobjective genetic algorithm (MOGA) and the Pareto curve to minimize the scaled function $\varphi_s + 1/\Delta P_s$, aiming to identify optimal values for OD and OF to maximize the mixing percentage ($\varphi$) and minimize the pressure drops ($\Delta$P). In contrast to neural networks, this study suggests employing Gaussian processes as a novel modeling technique. Gaussian processes offer the advantage of quantifying uncertainties, providing a range of acceptable values during modeling. So far, there has been no work in the field of microfluidics that has employed Gaussian processes as a modeling technique, making this approach an innovative contribution to the field. Additionally, the study utilized least-square optimization in conjunction with the MOGA to determine the optimal geometry.

In this study, we introduce the micromixer model, which is detailed in Section \ref{model}. Section \ref{results} presents the results and discussions, encompassing the stages of simulating data with the Gaussian process, optimizing parameters to determine the best outputs for the mixing percentage and pressure drop, and ultimately, comparing the resulting optimal values. The conclusion reflects on the achieved objectives and reinforces the use of GA to accelerate the optimization process.

\section{Methodology}\label{model}

\begin{figure}[h!]
\centering
\includegraphics[width=0.9\textwidth]{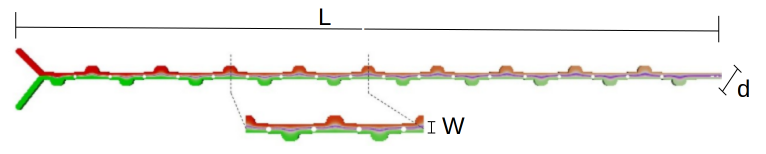}
\caption{Schematic diagram of the micromixer~\cite{maionchi2022}.}
\label{Fig2_modelo}
\end{figure}

This study is built upon the research presented in \cite{maionchi2022} and \cite{wang2012}, aiming to optimize the geometry proposed in \cite{maionchi2022}, shown in Figure~\ref{Fig2_modelo}. The dimension of the Y-geometry micromixer is described in Table~\ref{Tabela1}.  The optimization process involves comparing the conventional least-square method with a GA. To derive a reduced-order model with uncertainty quantification, Gaussian processes are employed. The objective is to determine the global minimum of the reduced-order model using the aforementioned optimization methods.

\begin{table}[h!]
\centering
\caption{Dimensions of micromixer.}
\begin{tabular}{l l}
\hline
Channel lenght (L) & $2cm$\\
Channel width (W) & $200\mu m$\\
Channel depth (d) & $20 \mu m$\\
Distance between two CGs & $1000 \mu m$\\
Diameter of CGs & $200 \mu m$\\
Distance between two obstructions & $1000 \mu m$\\
\hline
\end{tabular}
\label{Tabela1}
\end{table}

\subsection{Flow simulation}

The flow governing equations are given by
\begin{eqnarray}
     \overrightarrow{\nabla}.\overrightarrow{u} &=&0, \label{eq1_model}\\
     \rho \frac{D\overrightarrow{u}}{Dt}&=&- \overrightarrow{\nabla}p+\mu \nabla^{2} \overrightarrow{u},\label{eq2_model} \\
     \frac{DC}{Dt}&=&\gamma \nabla^{2} C,\label{eq3_model}
\end{eqnarray}
where $\overrightarrow {u}$, $p$ and $C$ are the velocity, pressure and concentration, respectively, and $\rho$, $\mu$ and $\gamma$ are the density, viscosity and fluid diffusion coefficient, respectively. The fluid is considered Newtonian  with constant properties and the flow uniform laminar. The fluid properties are density $\rho = 998kg/m^3$, dynamic viscosity $\mu =8.9 \cdot 10^{-4} Pa \cdot s$ and diffusion coefficient of the fluid $\gamma = 10^ {-9} m^2/s$.

The velocity of the flow is equal at both inputs, but the mass concentrations of solute are $C=1$ $mol/ m^3$ in the lower entry and $C = 0 \quad mol/ m^3$ in the upper entry (free of solute).
The velocity at the inputs is normal uniform, based on the Reynolds number, the walls are no slip and the gauge pressure at the output  is $p = 0 Pa$.
The Reynolds ($Re = \rho U_{in} W/\mu$) and Schmidt ($Sc = \mu/\rho \gamma$) numbers used in \cite{maionchi2022} are $Re \approx 1$ and $Sc = 890$, where $U_{in}$ is the flow velocity at the inputs and $W$ is the channel width.

The mixing percentage ($\varphi$) is determined by \cite{lin2007},

\begin{equation}
    \varphi = \left(1-\frac{\sigma}{\sigma_{max}} \right )\cdot 100\%, \label{phi}
\end{equation}
where $\sigma$ is the standard deviation, and the subscript \emph{max} denotes the initial unmixed state in the micromixer ($0.5$ in this case).
The standard deviation $\sigma$ can be calculated by

\begin{equation}
     \sigma = \sqrt{\frac{1}{N-1}\sum^{N}_{i=1}(C_{i}-\overline{C_{i}})^2}, \quad \overline{C_{i}} = \frac{\sum_{i=1}^N C_i}{N},
\end{equation}
where $N$ is the number of sampling across the channel width, $C_i$ is the concentration of sampling i, and $\overline{C_i}$ is the mean value of the concentration.
In addition, $\Delta P/ \varphi$ is also used to estimate the efficiency of the micromixers, as it measures the pumping power needed to obtain one percent of the mixture, where $\Delta P$ is the pressure difference between the output and inputs of the channel.

\subsection{Gaussian Process}

\begin{figure}[h!]
    \centering
    \includegraphics[width=\textwidth,trim=100pt 50pt 100pt 45pt, clip]{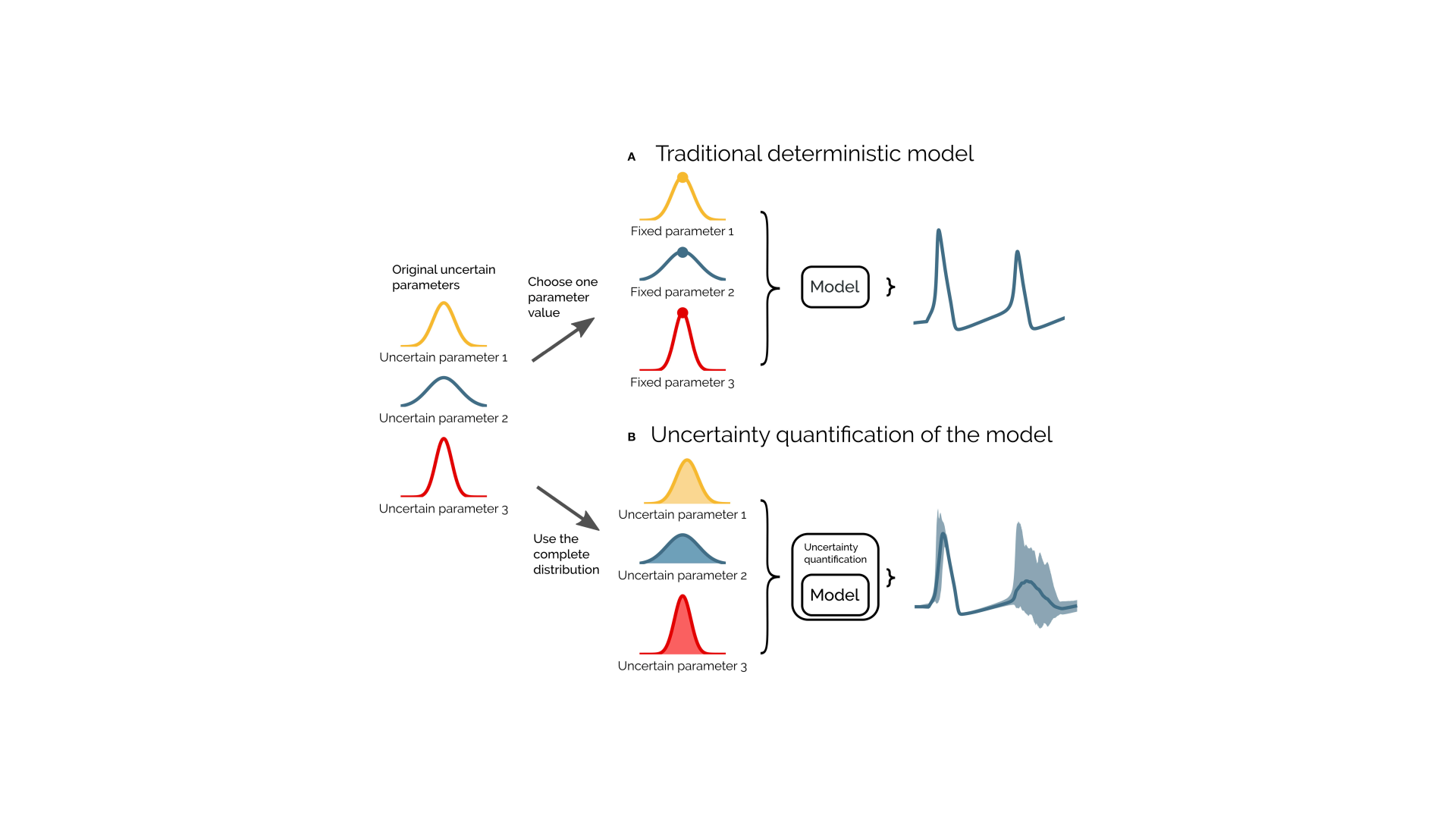}
    \caption{(A) In a traditional deterministic model each input parameter has a fixed value and there is a single outfput of the model. (B) In an uncertainty quantification of the model 
    each input parameter has a distribution and the output of the model becomes a range of possible values \cite{tennoe2018}.}
    \label{fig_model}
\end{figure}

In computational models, parameters are often uncertain, like variations in obstacle construction affecting their size and position in our study. While deterministic models assign single numerical values to parameters, as depicted in Figure 1, uncertainty quantification assigns distributions of possible values to each parameter. This approach, as shown in Figure \ref{fig_model}, propagates parameter uncertainty through the model, resulting in a distribution of model outputs. Sensitivity analysis, closely linked to uncertainty quantification, determines each parameter's impact on output uncertainty, as outlined by \cite{saltelli2002}. Parameters with high sensitivity cause significant output changes with small parameter variations, while those with low sensitivity lead to minor output variations.

The Gaussian processe (GP) is used as a data interpolator, which is entirely determined by its mean and covariance function.
This process, utilized in statistical modeling and Bayesian inference, is a stochastic process where the joint distributions of any finite subsets of random variables are Gaussian distributions. These multivariate Gaussian distributions are also able to specify the entire posterior predictive distribution, enabling the acquisition of uncertainty statistics.

The mean specifies the expected average value at any point in space or time, and the covariance function indicates how observations at different points are correlated with each other, according to
 
\begin{eqnarray}
    E[f(\bold{x})] &=& \mu(\bold{x}), \\
    Cov[f(\bold{x}),f(\bold{x}')] &=& k(\bold{x},\bold{x}') = E[(f(\bold{x})-\mu(\bold{x}))(f(\bold{x}')-\mu(\bold{x}'))],
\end{eqnarray}
where $E$ is the expected value, $\mu$ is the mean and, $Cov$ represents the covariance matrix and $k$ is the kernel function.

The multivariate Gaussian density, also known as marginal likelihood, implicitly integrates over all possible functions values at the set of all locations where the function is not observed.
Given a particular model, the marginal likelihood under GP prior of a set of function values $\{f(\bold{x}_1), f(\bold{x}_1), ... f(\bold{x}_N)\} = f(\bold{X})$ at $\bold{X}$ is givern by the following normal distribution \cite{rasmussen2006},

\begin{align}
    & p(f(\bold{X})|\bold{X},\mu,k) = \mathcal{N}(f(\bold{X})|\mu(\bold{X}),k(\bold{X},\bold{X})) \nonumber \\
    &= \frac{1}{\sqrt{(2\pi)^N |k(\bold{X},\bold{X})}|} \exp \left(-\frac{(f(\bold{x})-\mu(\bold{x}))^T k(\bold{X},\bold{X})^{-1} (f(\bold{X})-\mu(\bold{X}))}{2} \right).
\end{align}

Considering the observations, it is possible to say which function values are likely to be obtained at any new location. The predictive distribution of a function value $f(\bold{x}^*)$ at a test point $\bold{x}^*$ is given by \cite{rasmussen2006},
    
\begin{align}
    p(f(\bold{X}^*)|f(\bold{X}),\bold{X},\mu,k) = \mathcal{N}&(f(\bold{X}^*)|\mu(\bold{x^*})+k(\bold{x}^*,\bold{X})k(\bold{X},\bold{X})^{-1} (f(\bold{X})-\mu(\bold{X})), \nonumber \\
    & k(\bold{x}^*,\bold{x}^*) - k(\bold{x}^*,\bold{X}) k(\bold{X},\bold{X})^{-1} k(\bold{X},\bold{x}^*)),
\end{align}
where the predictive mean follows observations and the predictive variance shrinks given more data.

The kernel delineates the probable functions under the Gaussian Process (GP) prior, thus shaping the model's generalization properties. Each covariance function aligns with distinct assumptions regarding the function being modeled. Moreover, each kernel entails several parameters dictating the exact form of the covariance function. These parameters, often termed hyper-parameters, can be seen as defining a distribution over function parameters rather than directly specifying a function.

The distribution of the sum of GP functions with distinct kernels is also a GP function with the kernel given by 

\begin{equation}
    k(\bold{x},\bold{x}') = \sum_{type} \alpha_{type}^2  k_{type}(\bold{x},\bold{x}'),
\end{equation}
where each component can represent a different type of structure (for example, linear, period, squared-exp etc.).
The Radial Basis function (RBF) and the Rational Quadratic function (RQF), given by \cite{david2014}

\begin{align}
\label{kernel1}
    k_{RBF}(\bold{x},\bold{x}') &= \exp \left( -\frac{(\bold{x}-\bold{x}')^2}{2 l^2}\right), \\ 
\label{kernel2}    
    k_{RQF}(\bold{x},\bold{x}') &= \left(1 +\frac{(\bold{x}-\bold{x}')^2}{2 \alpha l^2}\right)^{-\alpha},
\end{align}
emerged as the most adept functions for the kernel adjustment in this study.

\section{Results}\label{results}

The ML model was trained considering the values of $OD$ and $OF$ as input variables and the values of $\Delta P$ and $\varphi$ as output.
Utilizing Gaussian processes with the kernels defined in Equations \ref{kernel1} and \ref{kernel2}, the following expression was derived to optimally fit the simulation data:

\begin{equation}
k = 1.71^2 k_{RBF}(l=0.652) + 0.387^2 k_{RQF}(\alpha=1.07e+05, l=0.331).\label{kernel}
\end{equation}

\begin{figure}[h!]
    \centering
    \includegraphics[width=0.49\textwidth,trim=550pt 50pt 600pt 150pt, clip]{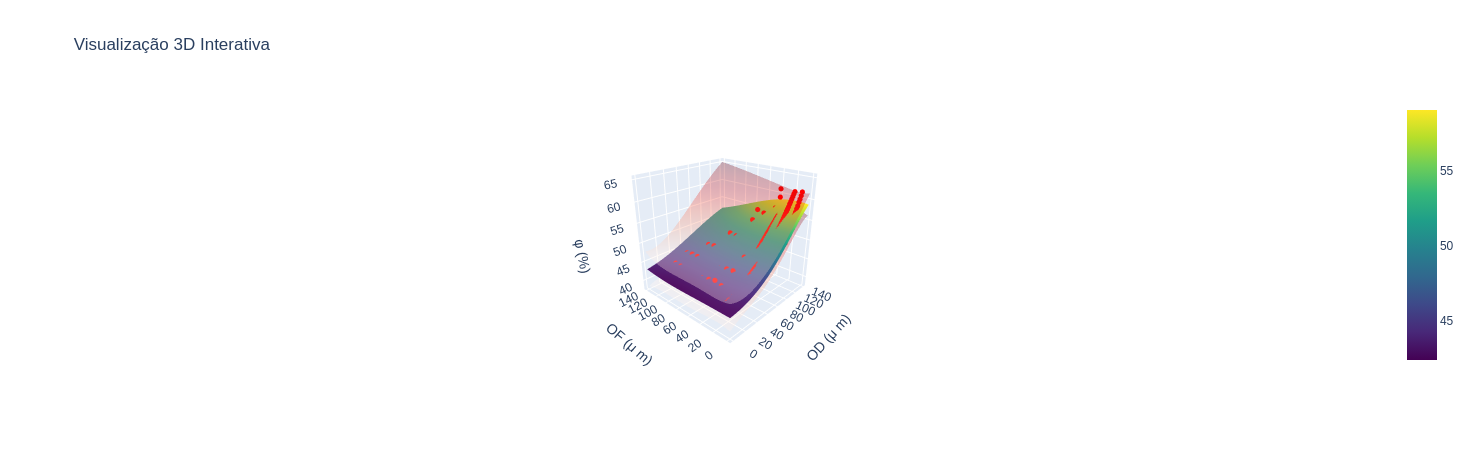}
    \includegraphics[width=0.49\textwidth,trim=550pt 50pt 600pt 150pt, clip]{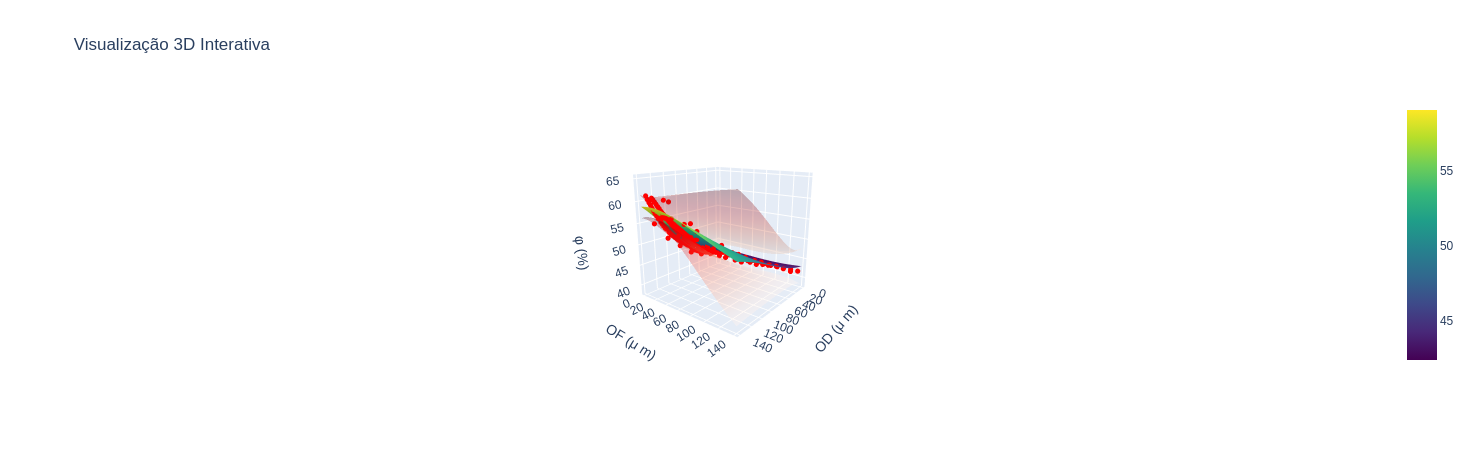}
    \includegraphics[width=0.49\textwidth,trim=550pt 50pt 600pt 150pt, clip]{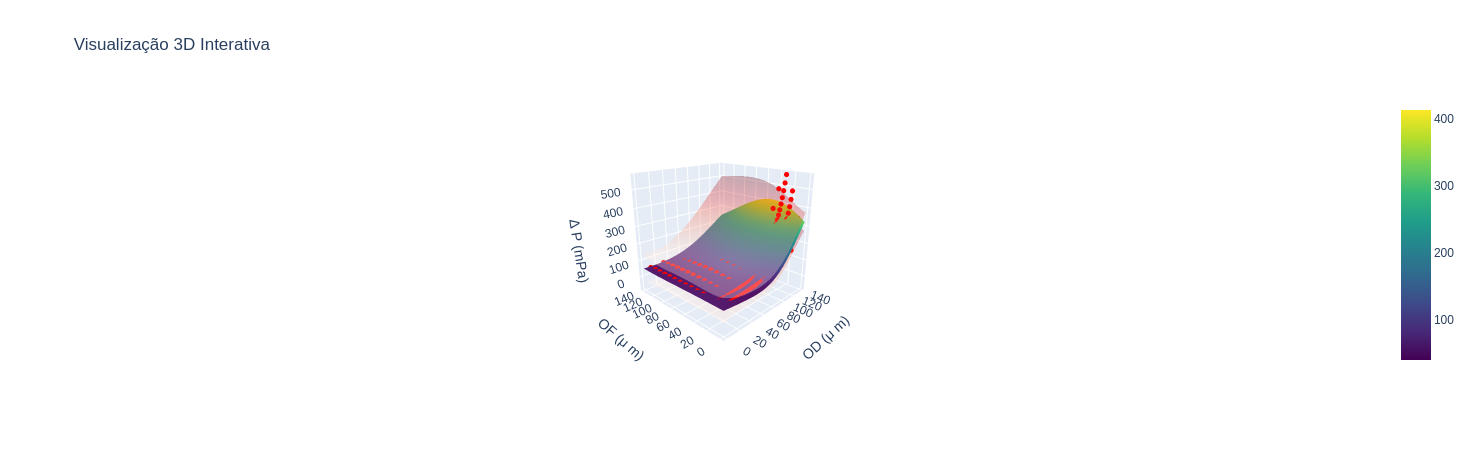}
    \includegraphics[width=0.49\textwidth,trim=550pt 50pt 600pt 150pt, clip]{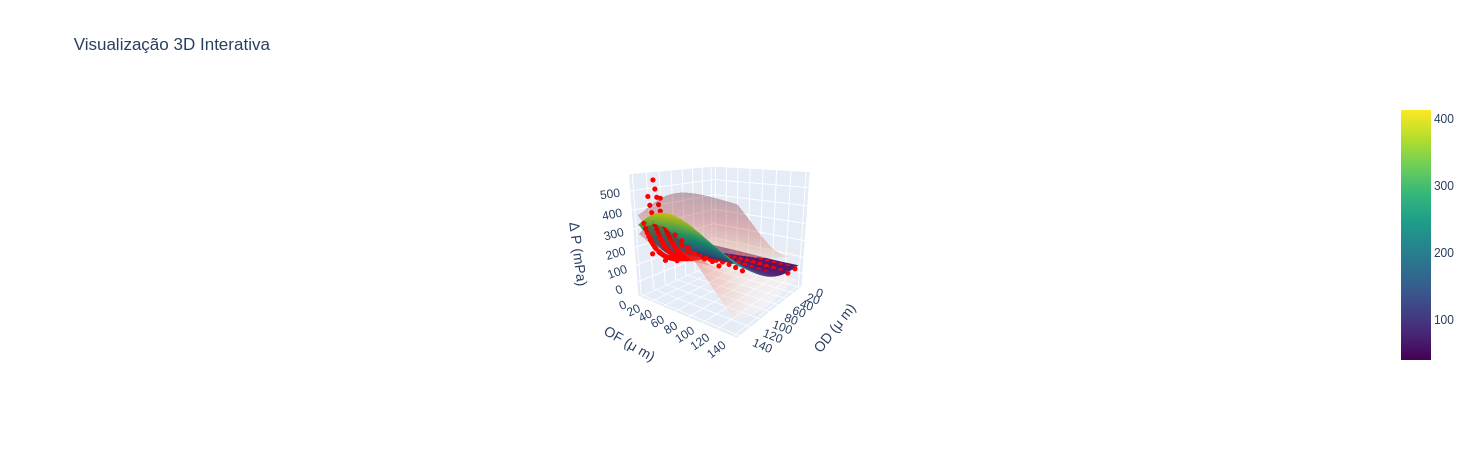}
    \caption{Simulation data in red depict mixing percentage (top) and pressure drop (bottom) as functions of geometric variables OD and OF. The plotted data are accompanied by the Gaussian process model, represented by the colorful surface, along with uncertainty bands in light red. The figures are presented from two different angles to enhance visualization and comparison of the data and the model.}
    \label{fig3D}
\end{figure}

Figure~\ref{fig3D} illustrates the simulation data of $\varphi$ and $\Delta P$ as functions of the geometric variables OD and OF, accompanied by the model obtained through the Gaussian process described in Eq.~\ref{kernel}, along with the associated uncertainty bands.
This three-dimensional view of the results reveals that the majority of the data falls within the range of uncertainty values, with this range being smaller for lower values of OD and OF.

\begin{figure}[t!]
    \centering
    \includegraphics[width=0.49\textwidth]{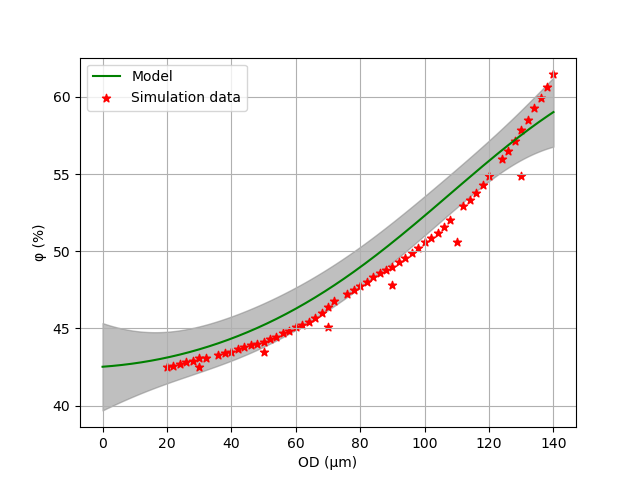}
    \includegraphics[width=0.49\textwidth]{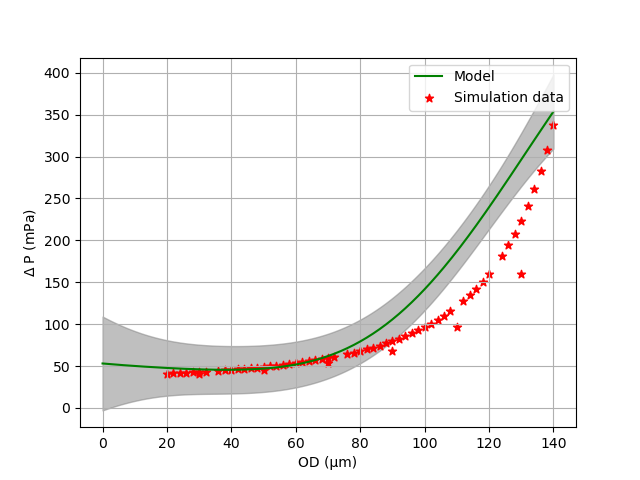}
    \includegraphics[width=0.49\textwidth]{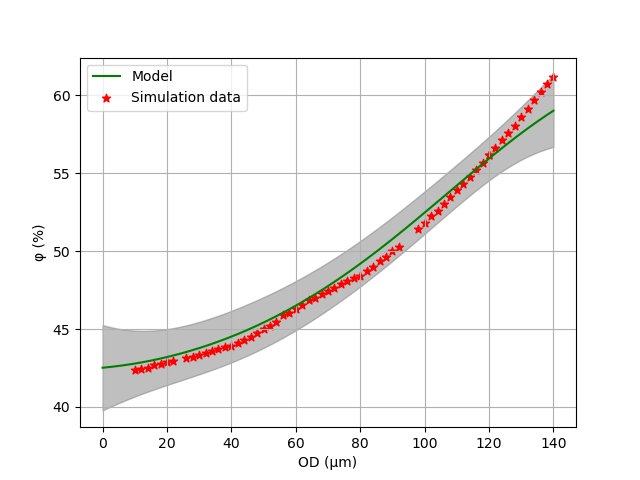}
    \includegraphics[width=0.49\textwidth]{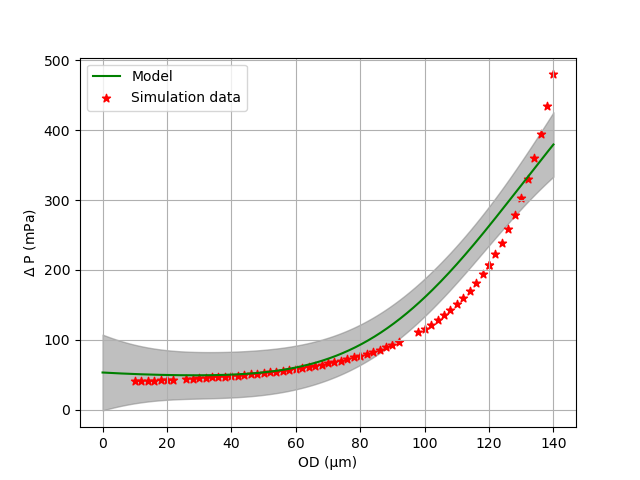}
    \caption{Projections of $\varphi$ (left) and $\Delta P$ (right) for fixed OF at values 10 and 20 $\mu$m.}\label{fig3}
\end{figure}

\begin{figure}[t!]
    \centering
    \includegraphics[width=0.49\textwidth]{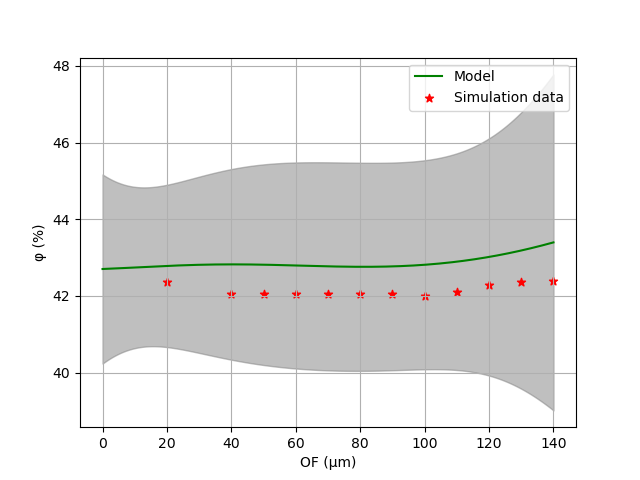}
    \includegraphics[width=0.49\textwidth]{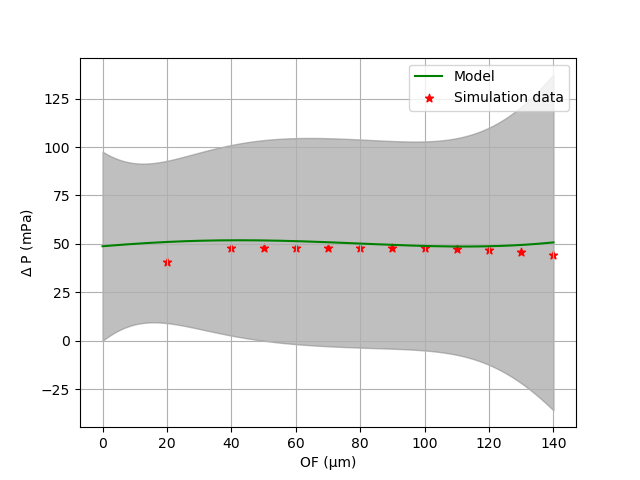}
    \includegraphics[width=0.49\textwidth]{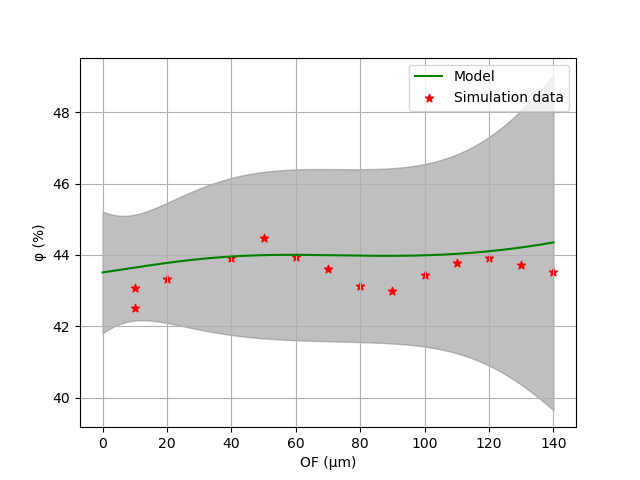}
    \includegraphics[width=0.49\textwidth]{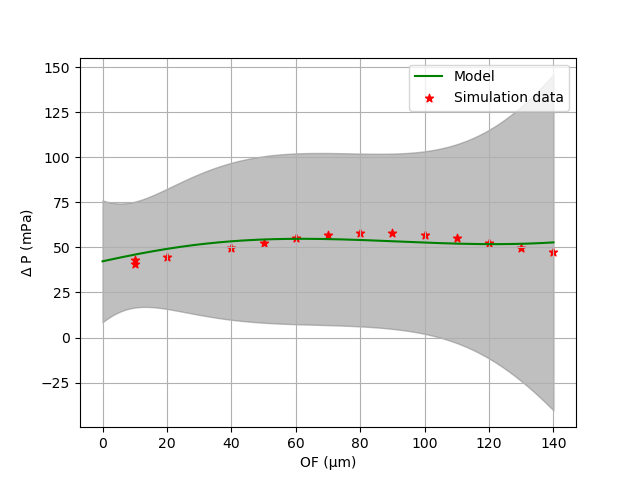}    
    \caption{Projections of $\varphi$ (left) and $\Delta P$ (right) for fixed OD at values 10 and 30 $\mu$m.}\label{fig4}
\end{figure}

Some projections of $\varphi$ and $\Delta P$ from Figure~\ref{fig3D} for fixed values of OD and OF are depicted in Figures~\ref{fig3} and \ref{fig4}. It is evident from both figures that lower values of OD and OF correspond to a greater amount of simulation data. 

In all cases, the reduced model closely aligns with the data. When OD is fixed, the uncertainty range is smaller compared to when OF is fixed. Although the uncertainty range tends to increase for higher values of OF, the data consistently adheres closely to the model, accurately capturing the rising profile with OD. Across different OD values, both the data and the model demonstrate nearly constant values across varying OF values, despite the relatively broad uncertainty range.

\begin{figure}[h!]
    \centering
    \includegraphics[width=0.8\textwidth]
    {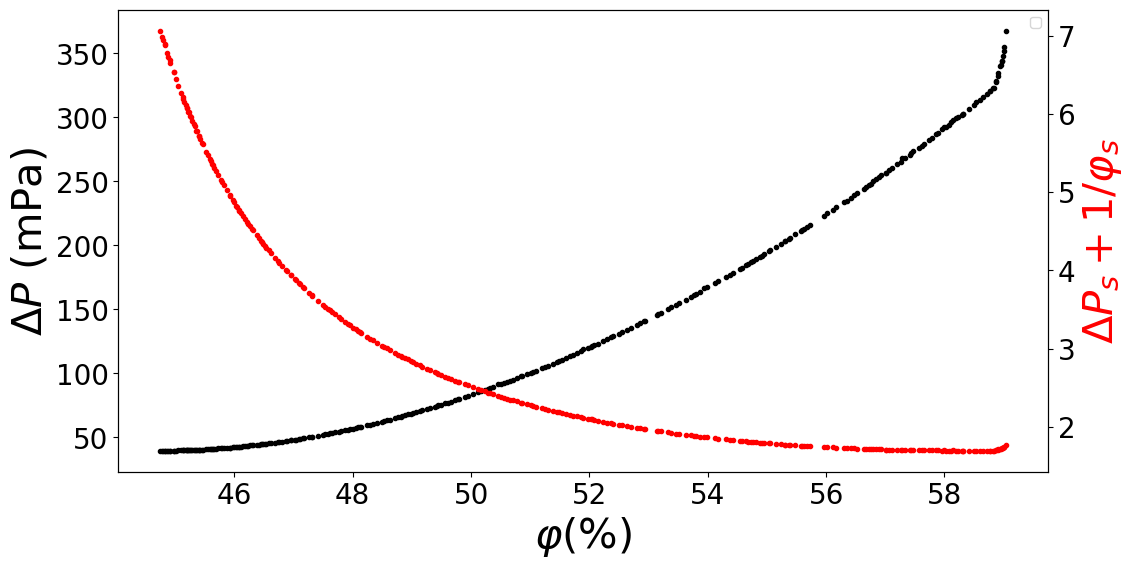}
    \includegraphics[width=0.8\textwidth]
    {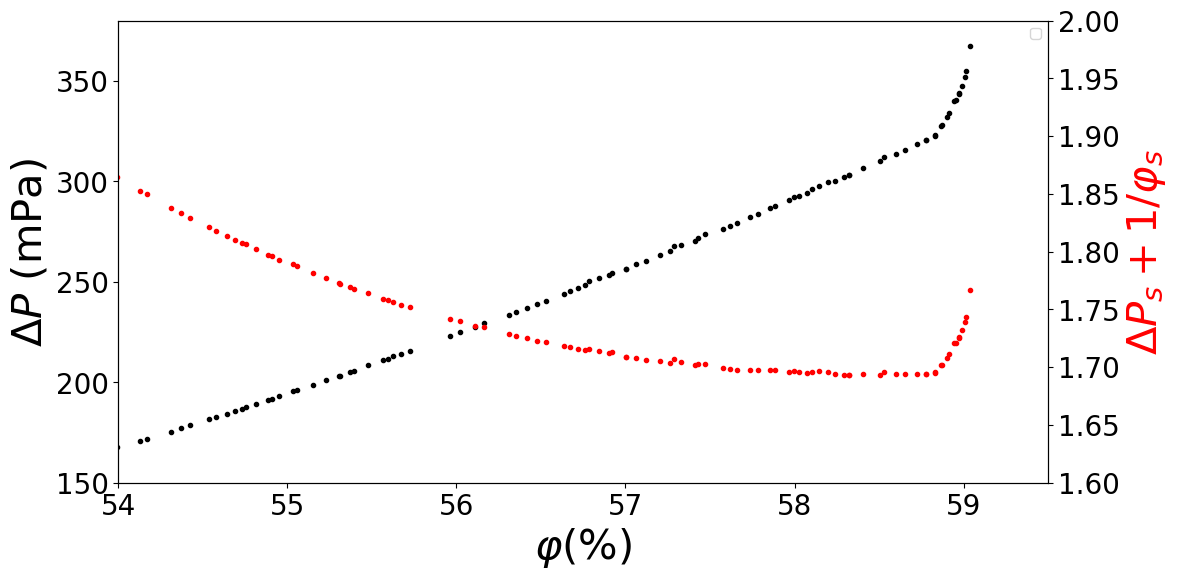}
    \caption{Pareto curves maximize $\varphi$ and minimize $\Delta P$ and $(\Delta P_s+1/\varphi_s)$ (top), and a zoomed-in view highlights the minimum of $(\Delta P_s+1/\varphi_s)$ (bottom). The results of the GA correspond to a population size of 300 with 5000 iterations.}
    \label{pareto}
\end{figure}

The optimization method employed in this study is based on a MOGA, specifically the NSGAII method, as described in a previous article \cite{maionchi2022}. This method aims to simultaneously satisfy the conditions of maximizing $\varphi$, minimizing $\Delta P$ and minimizing $(\Delta P_s+1/\varphi_s)$, where the variables in the last objective function are scaled according to their minimum and maximum values.
These objectives are achieved using Python's Platypus library \cite{platypus}. Additionally, the optimization process incorporates the same constraint condition utilized in the previous study, ensuring consistency. The input variables, objective functions, and constraints are explicitly defined to accurately address the problem. The resulting feasible solutions are then filtered from the set of all possible solutions. For ease of comparison, Pareto curves and simulation data are presented in Figure~\ref{pareto}.

From the simulation points, shown in black, it is noticeable that it is not possible to minimize $\Delta P$ and maximize $\varphi$ simultaneously, as both variables increase together. On the other hand, the Pareto curve, depicted in red, appears to exhibit the opposite behavior, where as $\Delta P$ decreases, $\varphi$ increases. Towards the end of the curve, a minimum point is reached, corresponding to the optimal situation sought after.
This particular case, along with its corresponding values of OD and OF, is shown in Table~\ref{Tabela3}. The optimization results using GA and the classical least-square method are also presented. The values of $\varphi$ and $\Delta P$ obtained in the previous work \cite{maionchi2022} are also listed in Table~\ref{Tabela3}, where it is noticeable that these values fall within the ranges calculated in this study. 

\begin{table}[h!]
\centering
\caption{Comparison between the values obtained with optimization and simulation for the optimal case.}
\begin{tabular}{l c c c c c c}
\hline
 Method & OD ($\mu$m) & OF ($\mu$m) &  $\varphi_{opt}$ & $\Delta P_{opt}$ (mPa) & $\varphi_{sim}$ & $\Delta P_{sim}$ (mPa)\\
\hline
 Article & $131$ & $10$ & $57.47$ & $235.87$ & $57.98$ &  $227.67$\\
\hline
 & &  & $55.96$ & $256.17$ & $$ &  $$\\
 GA & $137$ & $10$ & $58.32$ & $302.98$ & $60.32$ &  $295.31$\\
 & &  & $60.68$ & $349.80$ & $$ &  $$\\
 \hline
 & &  & $56.01$ & $258.73$ & $$ &  $$\\
 Least-Square & $137$ & $5$ & $58.41$ & $306.31$ & $61.47$ &  $302.82$\\
 & &  & $60.81$ & $353.88$ & $$ &  $$\\
\hline
\end{tabular}
\label{Tabela3}
\end{table} 

All predicted optimal outcomes of $\varphi$ and $\Delta$P are very similar, with corresponding OD varying from 131 to 137 $\mu$m and OF from 5 to 10 $\mu$m. These results are compared with the values obtained in the simulations. For the cases using GA and the earlier work with the neural network method, $\varphi_{sim}$ lies within the predicted range with GA; for the methods of GA and Least-square, both values of $\Delta P_{sim}$ lie within the predicted range of both methods. Thus, GA is the only method that had both simulation results lying within the predicted range. The corresponding geometry for all cases is positioned near the channel wall, featuring a medium-sized obstacle.

\begin{figure}[!b]
\begin{center}
\resizebox*{\textwidth}{!}{
\includegraphics[width=\textwidth,trim=30pt 300pt 30pt 300pt, clip]{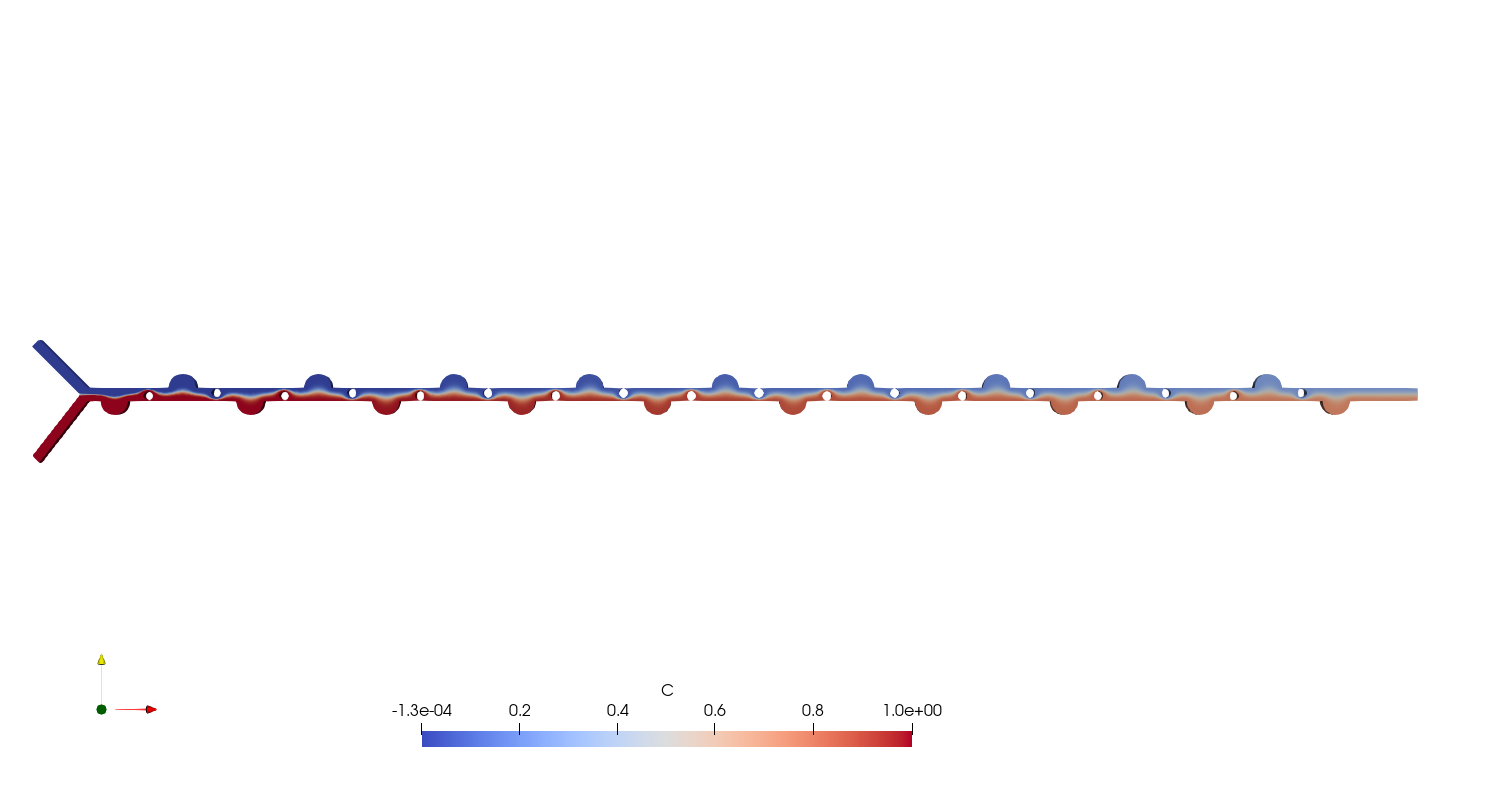}
}
\resizebox*{\textwidth}{!}{
\includegraphics[width=\textwidth,trim=0pt 300pt 0pt 200pt, clip]{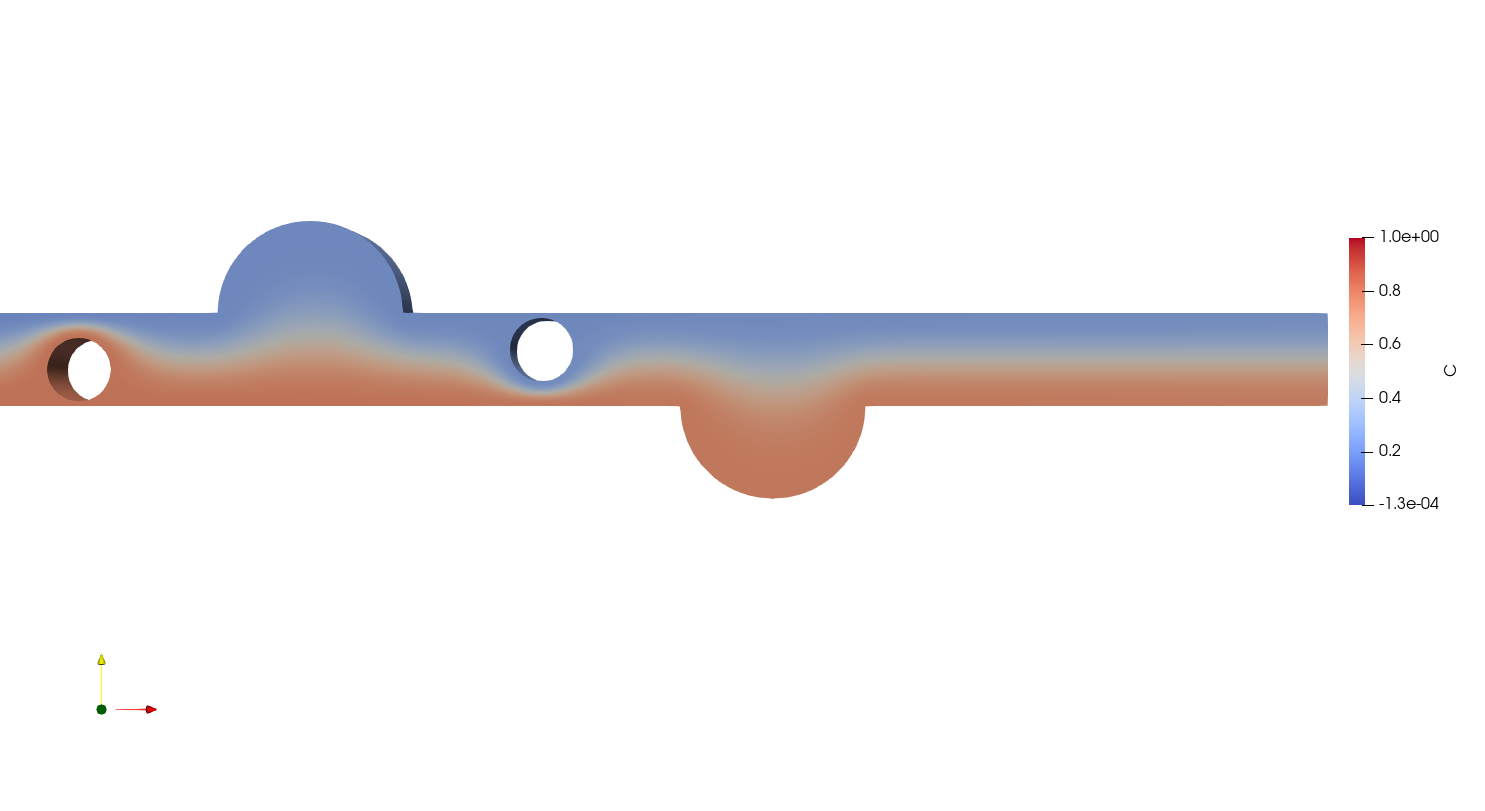}
}
\resizebox*{\textwidth}{!}{
\includegraphics[width=\textwidth,trim=30pt 300pt 30pt 300pt, clip]{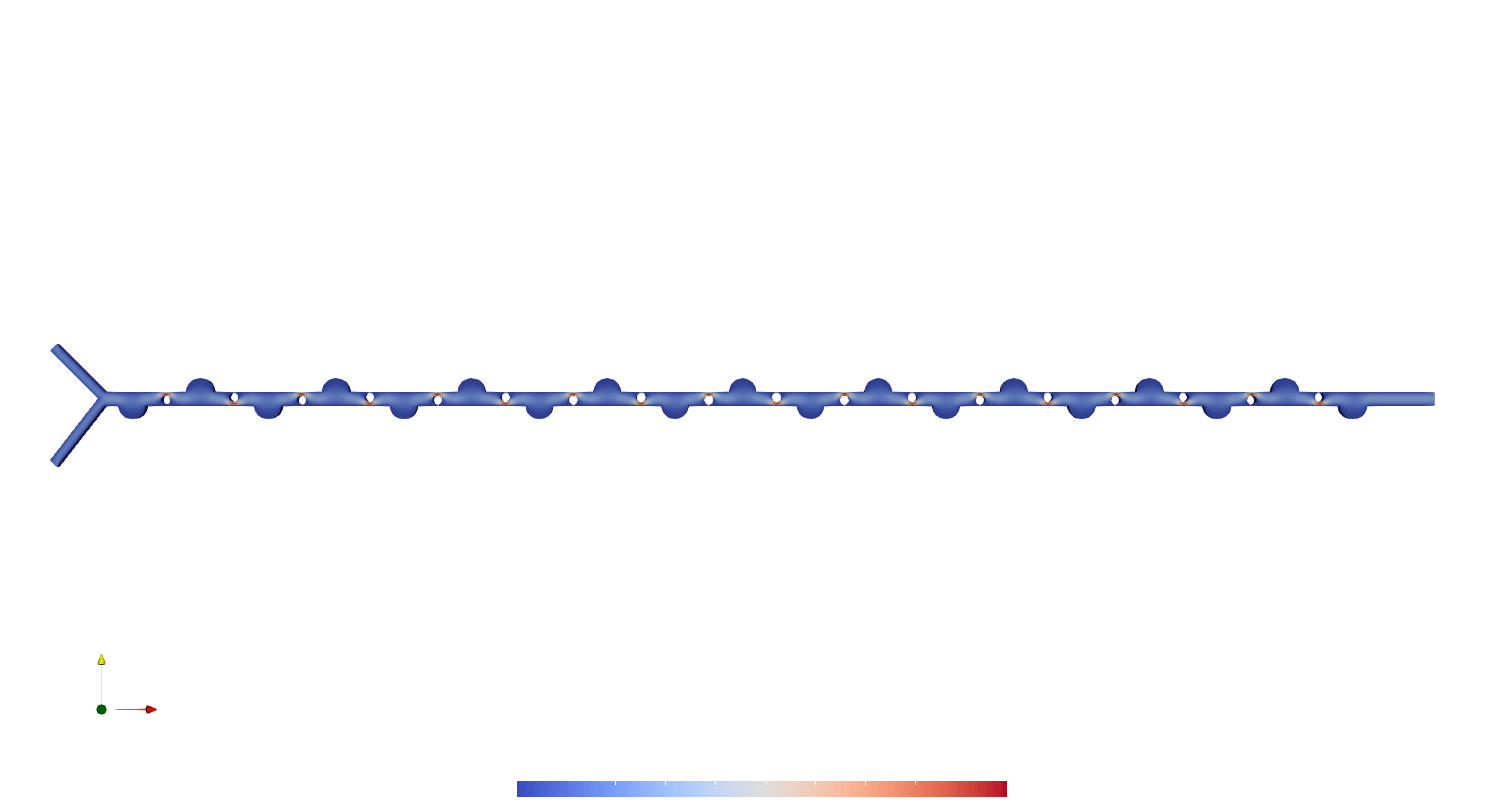}
}
\resizebox*{\textwidth}{!}{
\includegraphics[width=\textwidth,trim=0pt 300pt 0pt 200pt, clip]{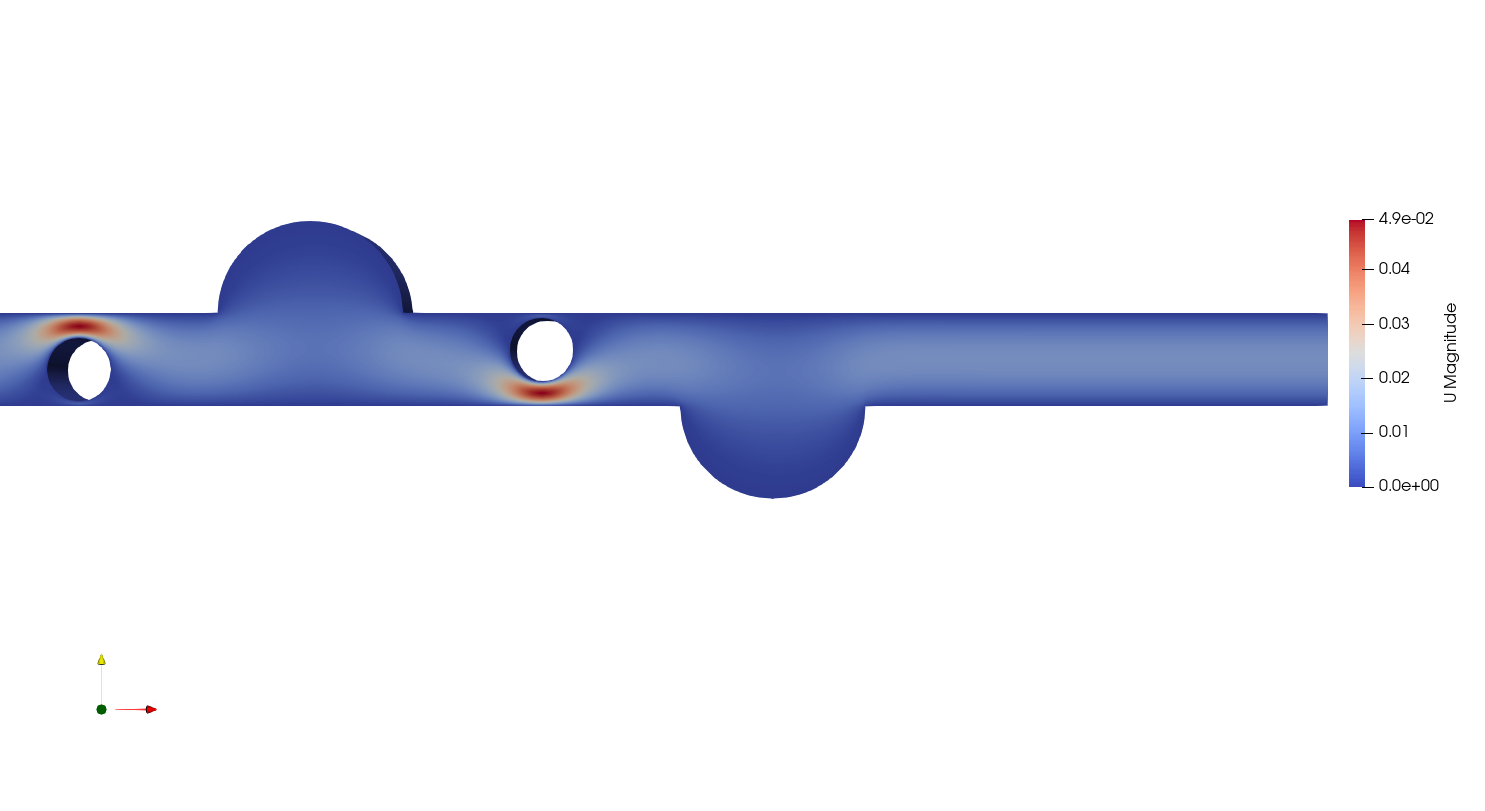}
}
\caption{Concentration ($mol/m^3$) (top) and velocity (m/s) (bottom) profiles of the optimal cases.}
\label{fig1_resultado}
\end{center}
\end{figure}

\begin{figure}[!t]
\begin{center}
\resizebox*{\textwidth}{!}{
\includegraphics[width=\textwidth,trim=30pt 300pt 30pt 300pt, clip]{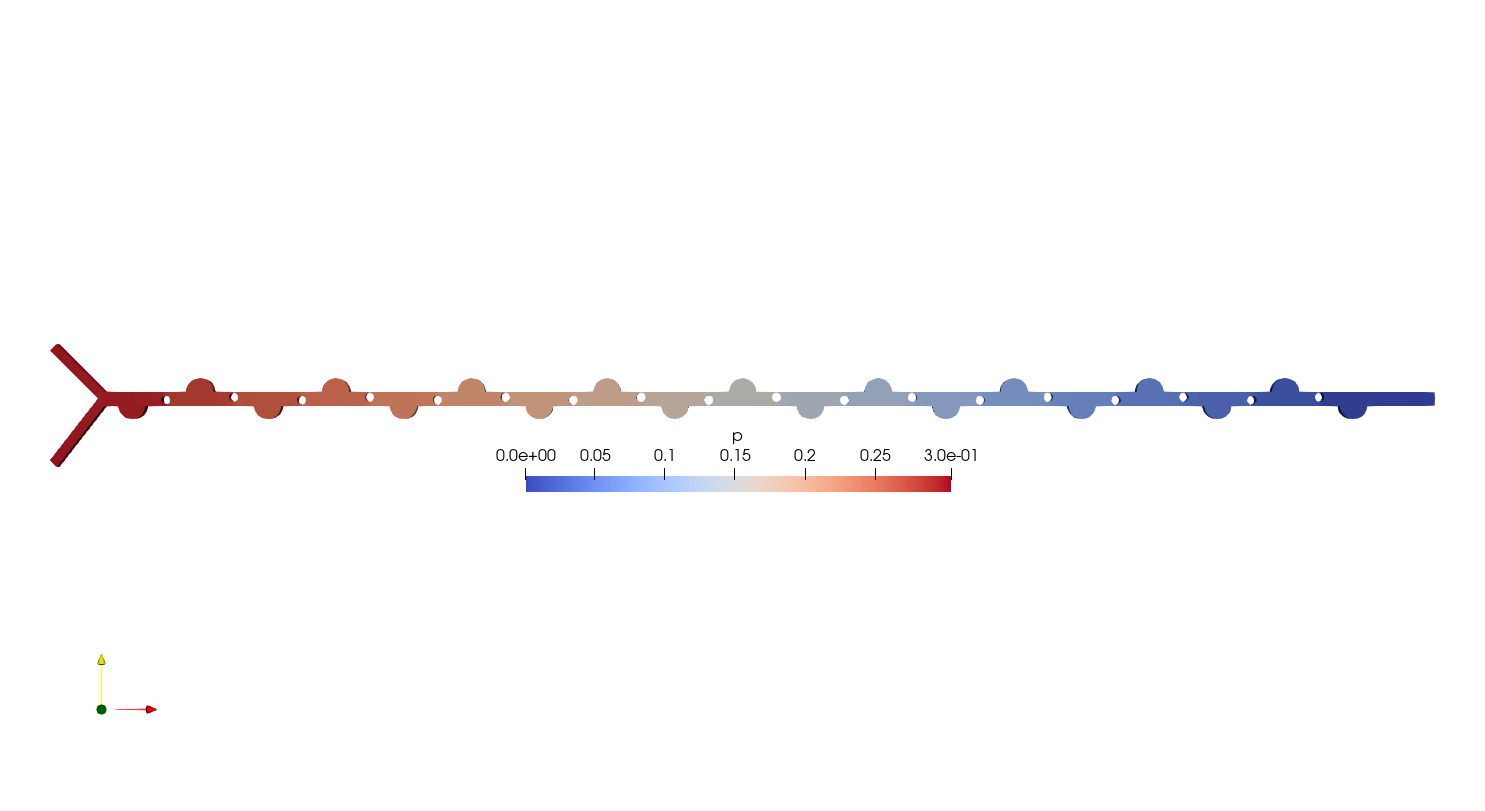}
}
\caption{Pressure (Pa) profile of the optimal case.}
\label{fig3_resultado}
\end{center}
\end{figure}

Figures~\ref{fig1_resultado} and \ref{fig3_resultado} depict the  concentration, velocity and pressure profiles obtained from the optimal data derived from the GA method. As already observed in the previous work \cite{maionchi2022}, the velocity field's asymmetry observed in Figure~\ref{fig1_resultado}, caused by the obstacles near the wall, induces changes in the concentration profile, leading to enhanced mixing. In Figure~\ref{fig3_resultado}, the pressure profile is evenly distributed due to the geometry's repetitive axial direction profile and symmetric inlet. 

\section{Conclusions}

Microfluidic devices have gained significant attention due to their compact size and efficient fluid handling capabilities. Micromixing, a crucial aspect in microfluidics, involves optimizing geometric parameters to enhance mixing efficiency. This study presents an innovative approach to microfluidic optimization by employing Gaussian processes with defined kernels to model simulation data.

Various simulations were conducted in a previous work, varying the diameter (OD) and offset (OF) of circular obstructions within the micromixer channel \cite{maionchi2022}. The obtained results were utilized to obtain Gaussian process model, which demonstrated high accuracy with uncertainty margins $13\%$ for mixing percentage ($\varphi$) and $15\%$ for pressure drop ($\Delta P$).

Global optimization based on the OD and OF parameters yielded an optimal micromixer geometry with circular obstacles, achieving an optimal configuration of OD = 137 $\mu$m and OF = 10 $\mu$m for $Re \approx 1$. These findings align with previous literature, highlighting the efficacy of the proposed methodology.

Verification through simulation revealed minimal errors for the optimized case, with $\varphi = 60.32 \%$ and $\Delta P = 295.31 $ mPa, demonstrating the robustness of the Gaussian process model. The utilization of Machine Learning techniques, specifically GAs, proved effective in addressing optimization challenges in microfluidic systems.

In conclusion, this study validates the CFD model, demonstrates the effectiveness of Gaussian processes in capturing system non-linearity, and establishes it as a viable surrogate model for global optimization in microfluidic systems. The proposed methodology significantly accelerates the optimization process, providing valuable insights for the design and optimization of micromixer devices.

\bibliographystyle{elsarticle-num}
\bibliography{ref}

\end{document}